\documentclass[pre,twocolumn,showpacs]{revtex4}

\usepackage{amsfonts}
\usepackage{epsfig}
\usepackage{graphicx}
\usepackage{amsmath}
\usepackage{mathrsfs}
\usepackage{dcolumn}
\newcommand{\beq}{\begin{eqnarray}}
\newcommand{\eeq}{\end{eqnarray}}

\newcommand{\be}{\begin{equation}}
\newcommand{\ee}{\end{equation}}
\newcommand{\bey}{\begin{eqnarray}}
\newcommand{\eey}{\end{eqnarray}}
\newcommand{\ba}{\begin{array}}
\newcommand{\ea}{\end{array}}
\newcommand{\bi}{\begin{itemize}}
\newcommand{\ei}{\end{itemize}}
\newcommand{\bem}{\begin{enumerate}}
\newcommand{\eem}{\end{enumerate}}
\newcommand{\bw}{\begin{widetext}}
\newcommand{\ew}{\end{widetext}}
\newcommand{\ra}{\rangle}
\newcommand{\la}{\langle}

\newcommand{\ww}{\widetilde}
\newcommand{\WW}{\widetilde}

\newcommand{\E}{{\cal E}}
\newcommand{\T}{{\cal T}}

\newcommand{\N}{{\cal{K} }}

\newcommand{\HH}{{\mathcal{H}}}
\newcommand{\tr}{{\rm Tr}}

\begin{document}

 \title{Statistical description of small quantum systems beyond weak-coupling limit}

\author{Wen-ge Wang}
\affiliation{
 Department of Modern Physics, University of Science and Technology of China,
 Hefei 230026, China }
\email{wgwang@ustc.edu.cn}
 \date{\today}

 \begin{abstract}

 An explicit expression is derived for the statistical description of small quantum systems,
 which are relatively-weakly and directly coupled to only small parts of their environments.
 The derived expression has a canonical form, but is given
 by a renormalized self-Hamiltonian of the studied system, which appropriately takes
 into account the influence of the system-environment interaction.
 In the case that the system has a narrow spectrum and the environment is
 sufficiently large, the modification to the self-Hamiltonian usually has a
 mean-field feature, given by an environmental average of the interaction Hamiltonian.
 In other cases, the modification may be beyond the mean-field approximation.

 \end{abstract}
%05.30.-d Quantum statistical mechanics
%03.65.-w   Quantum mechanics
%03.65.Yz   Decoherence; open systems; quantum statistical methods

 \maketitle

%\tabSPofcontents

 \section{Introduction}

 Statistical description of small quantum systems, which have non-very-weak interaction with environments,
 is of interest in a variety of fields, ranging from the field of ultra-cold atoms, to chemical physics,
 condensed-matter physics, bio-physics, and so on.
 Recently, development of technologies has made it possible to perform direct experimental studies on
 thermalization of systems of the mesoscopic, even microscopic scale \cite{Weiss06,Hoff07},
 showing deviation from the standard canonical distribution due to non-weak interaction \cite{Weiss06}.
 Meanwhile, theoretical studies and numerical simulations have been also carried out, some
 showing approach to ordinary thermal states \cite{Rigol08,DSFVJ12},
 while some suggesting significant deviations
 in certain cases \cite{Kollath07,Manma07,EK08,Haangi08,BKL10,BCH11,GME11}.

 Although known for a long time that
 an equilibrium state of a quantum system can be described by a canonical distribution
 when the system-environment interaction is sufficiently weak \cite{Schrodinger52,LL-SP},
 a sound foundation has been established only recently \cite{Tak98,Gold06,Pepes06,RGE12,GMM09,DYLS07,
 LPSW09,BG09,GLMTZ10,Kim10}.
 A further question arises for small quantum systems, whose interaction with environments is
 usually non-negligible:
 In which way does the system-environment interaction influence the statistical description of
 such a small quantum system?
 Up to now, an explicit and general expression for the statistical description of such systems has
 not been derived, yet, from the microcanonical-ensemble description of total systems.

 As an example, one may consider a spin at a given site, whose interaction with
 surrounding spins as the environment is described by an interaction Hamiltonian
 $H^I = \epsilon J^S J^A$, where $\epsilon$ is a parameter characterizing the strength of the interaction
 and $J^S$ and $J^A$ are operators acting on the Hilbert spaces of the system and the environment,
 respectively.
 The problem we are to address is:
 When $\epsilon$ is not very small, i.e., beyond the regime of
 weak-coupling limit in which the canonical distribution has been derived \cite{Gold06,Pepes06,RGE12},
 does the spin still have a  statistical description of the canonical form and, if yes,
 in which way does the interaction term contribute to the description?

 Interestingly, a similar situation is met in the field of decoherence when studying the so-called
 preferred (pointer) basis, in which reduced density matrices become approximately diagonal
 with time passing.
 The preferred basis has been known to be close to the energy eigenbasis of the system in the case of
 sufficiently-weak system-environment interaction \cite{Zurek03,WGCL08,Gogolin10}.
 While, recently, considerable deviation has been observed due to non-weak interaction \cite{WHG12}.

 In this paper, it is shown that, taking advantage of techniques developed in the canonical-typicality
 approach to the foundations of statistical mechanics \cite{Tak98,Gold06,Pepes06,RGE12,GMM09,DYLS07,
 LPSW09,BG09,GLMTZ10,Kim10}, an explicit expression can be derived for the statistical
 description of small systems, which have weak but not very weak interaction with environments.
 The derived expression still has a canonical form, but is given
 by a renormalized self-Hamiltonian of the studied system, which appropriately takes
 into account the influence of the system-environment interaction and reduces to the ordinary
 self-Hamiltonian in the weak-coupling limit.
 In some cases, the expression may hold even for non-weak interaction.

 \section{Main result}
 \label{sect-mainr}

 In this section, we present our main result and discuss its physical meaning.
 (The proof will be given in the next section.)
 Before doing this, below we first introduce the settings and notations to be used in presenting
 the main result.

 We'll consider a total system denoted by $\T$, which is composed of a small system $S$ of interest and
 an environment $\E$ much larger  than $S$, under a total Hamiltonian,
 \begin{equation}\label{}
 H = H^{S} + H^{I} + H^{\E},
 \end{equation}
 where $H^S$ and $H^\E$ are the Hamiltonians of $S$ and $\E$ determined in the weak-coupling
 limit and $H^I$ represents the interaction.
 A parameter $\epsilon$ is used to characterize the strength of the interaction, with
 $H^I \propto \epsilon$.
 Normalized eigenstates of $H$ are denoted by $|E_\eta\ra$ with energies $E_\eta$,
 those of $H^S$ denoted by $|E^S_\alpha\ra $ with energies $E^S_\alpha$, and
 those of $H^\E$ denoted by $|E^{\E}_i\ra $ with energies $E^\E_i$,
 namely,
 \begin{subequations}
\begin{eqnarray}
  H |E_\eta\ra = E_\eta |E_\eta\ra,
 \\  H^S |E^S_\alpha\ra = E^S_\alpha |E^S_\alpha\ra ,
 \\  H^\E |E^{\E}_i\ra = E^{\E}_i |E^{\E}_i\ra .
\end{eqnarray}
 \end{subequations}
 We use $\HH_{\delta E}$ to denote a subspace of the total Hilbert space, which is spanned by
 those eigenstates $|E_\eta\ra$ with energies $E_\eta$ lying in a narrow energy region $[E,E+\delta E]$.
 Here, $\delta E$ is small but not too small, such that there are sufficiently many
 energy levels within $[E,E+\delta E]$ for statistical treatment.
 For the simplicity in discussion, we assume that all the Hamiltonians discussed have
 non-degenerate spectra.

 We'll consider a renormalized  self-Hamiltonian of the system $S$, as well as
 a renormalized interaction Hamiltonian, denoted by $\ww H^S$ and $\ww H^I$, respectively,
 which satisfy the following relation,
 \begin{equation}
 \label{reHS} \ww H^{S} = H^{S} + H^I_S, \ \ \ww H^{I} = H^{I} - H^I_S.
 \end{equation}
 Note that the total Hamiltonian $H$ remains unchanged.
 Here, the term $H^I_S$, representing certain impact of $H^I$ in the system $S$,
 is an operator acting on the Hilbert space of $S$.
 We use $|E_{\ww\alpha}\ra$ to denote eigenstates of the renormalized
 self-Hamiltonian $\ww H^S$ with eigenenergies $E_{\ww\alpha}$,
 \begin{equation}\label{}
  \ww H^{S} |E_{\ww\alpha}\ra = E_{\ww\alpha} |E_{\ww\alpha}\ra .
 \end{equation}
 (Note that, for brevity, we omit a subscript $S$ in the notation $|E_{\ww\alpha}\ra$.)

 The operator $H^I_S$ is defined by the following relation in the renormalized basis, namely,
 \begin{equation}\label{HIS}
 (H^I_S)_{\ww\alpha \ww\beta} = \la H^{I}_{\ww \alpha \ww\beta} \ra_{\ww\gamma} ,
 \end{equation}
 where $(H^I_S)_{\ww\alpha \ww\beta} \equiv \la E_{\ww\alpha}|H^I_S |E_{\ww\beta}\ra $.
 Here, $H^{I}_{\ww\alpha \ww\beta} \equiv \la E_{\ww\alpha}|H^I |E_{\ww\beta}\ra$ is an
 operator acting on the Hilbert space of the environment.
 We use $\HH^{(\E)}_{\ww\gamma}$ to denote a subspace in the Hilbert space of the environment $\E$,
 which is spanned by those states $|E^{\E}_i\ra $ with energies
 $E^{\E}_i \in [E-E_{\ww\gamma},E-E_{\ww\gamma} +\delta E]$.
 The right hand side of Eq.(\ref{HIS}), namely, $\la H^{I}_{\ww\alpha \ww\beta}\ra_{\ww\gamma} $,
 represents the average of the expectation value $\la E^{\E}_i|H^{I}_{\ww\alpha \ww\beta}|E^{\E}_i\ra $
 over $|E^{\E}_i\ra  \in \HH^{(\E)}_{\ww\gamma}$, where
 $\ww \gamma  = \ww \beta$ if $E_{\ww \alpha} > E_{\ww \beta}$ and $\ww\gamma =\ww\alpha$
 if $E^S_{ \ww\alpha} < E^S_{ \ww\beta}$ \cite{foot-HIS}.

 We use $\rho^S$ to denote the reduced density operator of the system $S$, which is
 obtained from the microcanonical-ensemble description of the total system in the subspace
 $\HH_{\delta E}$, by tracing out the degrees of freedom of the environment.
 That is,
 \begin{equation}\label{}
 \rho^S = \frac{1}{N_{\delta E}}
 \tr_\E \left ( \sum_{|E_\eta\ra \in \HH_{\delta E}} |E_\eta\ra \la E_\eta | \right ) ,
 \end{equation}
 where $N_{\delta E}$ is the dimension of the subspace $\HH_{\delta E}$,
 \begin{equation}\label{}
 N_{\delta E} = \sum_{|E_\eta\ra \in \HH_{\delta E}} 1.
 \end{equation}

 The main result of this paper is that, under conditions specified below, $\rho^S$ has
 the following canonical form of expression with parameter $\beta_T$,
 \be \label{rho-c} \rho^S  \simeq  e^{- \beta_T \WW H^S }/\tr e^{- \beta_T \WW H^S}.
 \ee
 The conditions are:
 (i) The environment $\E$ can be divided into two parts $A$ and $B$, $\E = A + B$,
 such that the interaction Hamiltonian $H^{I}$ couples $S$ to $A$ only,
 where $A$ is a small system and $B$ is much larger than $A$.
 (ii) The environment $\E$ has certain type of complexity.
 (iii) The system-environment interaction is relatively weak.
 (iv) The width $\delta E$ is fixed, or decreases sufficiently slowly with increasing size of
 the environment.
 Details of the last three conditions will be given in the next section.

 To see the physical meaning of the renormalized self-Hamiltonian $\WW H^S$,
 let us consider the example of spin mentioned above.
 For $H^{I} = \epsilon J^S J^A$, one has $ \la H^{I}_{\ww\alpha \ww\beta} \ra_{\ww\gamma}
 = \epsilon J^S_{\ww\alpha \ww\beta} \la J^A \ra_{\ww\gamma} $,
 where  $J^S_{\ww\alpha \ww\beta} = \la E_{\ww\alpha} |J^S|E_{\ww\beta}\ra$ and
 $\la J^A \ra_{\ww\gamma}$ is the average of the expectation value of $J^A$ in the subspace
 $\HH^{(\E)}_{\ww\gamma}$.
 Then, making use of Eq.(\ref{HIS}), the renormalized self-Hamiltonian can be written as
 \begin{eqnarray}\label{HSfull}
 \WW H^S = H^{S} +\epsilon \overline{J^A} J^S  + \Delta H^{S},
 \end{eqnarray}
 where $\overline{J^A}$ is the average of $\la J^A \ra_{\ww\gamma}$ over $\ww\gamma $ and
 \begin{eqnarray} \label{DHS}
  \Delta H^{S}= \epsilon \sum_{\ww\alpha \ww\beta}
 (\Delta J^A_{\ww\alpha \ww\beta}) J^S_{\ww\alpha \ww\beta} \ |E_{\ww\alpha}\ra \la E_{\ww\beta} |,
 \end{eqnarray}
 with $\Delta J^A_{\ww\alpha \ww\beta} = \la J^A \ra_{\ww\gamma} - \overline{J^A} $.
 The term $\epsilon \overline{J^A} J^S$ can be interpreted as
 interaction of the system $S$ with a mean field $\overline{J^A}$.

 In the case that $\la J^A \ra_{\ww\gamma}$ is almost the same for all the subspaces
 $\HH^{(\E)}_{\ww\gamma}$,
 $\Delta J^A_{\alpha \beta} \simeq 0$ and $\WW H^S \simeq H^{S} + \epsilon \overline{J^A} J^S$.
 For example, this is the case, if the environment $\E$ is sufficiently large compared with the system $S$
 and a change in the energy of $S$ has negligible influence in properties of $\E$.
 On the other hand, if variation of $\la J^A \ra_{\ww\gamma}$ with ${\ww\gamma}$ is not small,
 $\Delta J^A_{\alpha \beta}$ can not be neglected; in this case,
 modification to the self-Hamiltonian is beyond the mean-field approximation.
 (Generalization of the above discussions to the general case of $H^{I} = \epsilon \sum_l J^S_l J^A_l$
 is straightforward.)

 \section{Proof of the main result}

 In this section, we give our proof for the above discussed main result.
 First, we specify the second condition mentioned in the previous section below Eq.(\ref{rho-c}),
 about complexity of $\E$.
 For the simplicity in discussion, at this stage we consider a strong version of the second
 condition, that is,  we assume that the environment is
 a quantum chaotic system that can be described by the random matrix theory \cite{Sredn94}.
 (Later, at the end of Sec.\ref{sect-swi}, we'll discuss to what extent this restriction to $\E$
 can be loosed for the purpose here and propose a weak version of the second condition.)
 Specifically, in the expansion
 \begin{equation}\label{Ei-expan}
 |E^\E_i\ra = \sum_{m, q} C^i_{mq} |m_A\ra |q_B\ra ,
 \end{equation}
 where $|m_A\ra $ indicate an orthonormal basis in the Hilbert space of the system $A$ and $|q_B\ra $ for
 an orthonormal basis in that of $B$, and the real and imaginary parts of the components $C^i_{mq}$
 can be regarded as Gaussian random numbers with mean zero and the same variance.
 A related quantity that will be used is
 \begin{equation}\label{G}
 G^{ij}_{mm'} \equiv  \la E^\E_{i} |m_A \ra \la m_A'|E^\E_{j} \ra .
 \end{equation}

 Next, we introduce other notations to be used in the proof.
 We use $N_S$, $N_A$, $N_B$, and $N_\E$ to denote the dimensions of the Hilbert spaces of the systems
 $S$, $A$, $B$, and $\E$, respectively.
 We use $\rho^S_{\delta E}$ to denote the reduced density matrix
 given by a typical vector $|\Psi_{\delta E}^{\rm ty}\ra \in \HH_{\delta E}$, namely,
 \be \rho^S_{\delta E} = \tr_{\E} |\Psi_{\delta E}^{\rm ty}\ra \la \Psi_{\delta E}^{\rm ty}|.
 \ee
 Similar to $\HH^{(\E)}_{\ww \gamma}$ defined in the previous section,  we use
 $\HH^{(\E)}_\alpha$ to denote the subspace spanned by those states $|E^{\E}_i\ra $ with energies
 $E^{\E}_i \in [E-E^{S}_\alpha,E-E^{S}_\alpha +\delta E]$ in the Hilbert space of the environment $\E$.
 Dimension of the subspace $\HH^{(\E)}_\alpha$ is denoted by $N^{(\E)}_\alpha$.
 An important subspace to be considered below is $\HH_d$, defined by
 \be \label{HHd} \HH_{d} = \bigoplus_{\alpha} |E^S_\alpha\ra \otimes \HH^{(\E)}_\alpha.
% \ \ \ww \HH_{d} = \bigoplus_{\ww \alpha} |E_{\ww \alpha}\ra \otimes \HH^{(\E)}_{\ww\alpha}.
 \ee
 We use  $\rho^S_d$ to denote the reduced density operator given by
 a typical vector $|\Psi_{d}^{\rm ty}\ra \in \HH_d$, namely,
 \be \rho^S_{d} = \tr_{\E} |\Psi_{d}^{\rm ty}\ra \la \Psi_{d}^{\rm ty}|.
 \ee

 We'll consider elements of the total Hamiltonian $H$.
 In the basis $|E^{S}_\alpha E^\E_i\ra$, the elements are denoted by
 \be H_{\alpha i,\beta j} \equiv \la E^S_\alpha E^\E_{i}|H |E^S_\beta E^\E_{j}\ra.
 \ee
 Similarly, for the interaction Hamiltonian,
 \be H^I_{\alpha i,\beta j} \equiv \la E^S_\alpha E^\E_{i}|H^I |E^S_\beta E^\E_{j}\ra .
 \ee
 Like $H^I_{\ww\alpha \ww\beta}$ introduced previously, we define
 \begin{equation}\label{Hab}
 H^I_{\alpha\beta} \equiv \la E^S_\alpha| H^I |E^S_\beta\ra, \ \ \
 \WW H^I_{\ww\alpha \ww\beta} \equiv \la E_{\ww\alpha} |\WW H^I|E_{\ww\beta} \ra ,
 \end{equation}
 which are in fact operators acting on the Hilbert space of the small system $A$.
 Correspondingly, we have elements
 $(H^I_{\alpha\beta})_{ij} \equiv \la E^\E_i| H^I_{\alpha\beta}|E^\E_j\ra$
 and $(\WW H^I_{\ww\alpha \ww\beta})_{ij} \equiv \la E^\E_i|\WW H^I_{\ww\alpha \ww\beta}| E^\E_j\ra$,
 and it is easy to see that
 \be \label{HIab2} H^I_{\alpha i,\beta j} = (H^I_{\alpha\beta})_{ij}.
 \ee

 The following two quantities will also be used:
 One is the following mean value,
 \begin{equation}\label{hdia}
 h^{\rm dia}_{\alpha \beta} \equiv \tr_A (H^{I}_{\alpha \beta})/N_A,
 \end{equation}
 the other, indicated by $h$, is the maximum of the mean value
 $(\sum_{m m'} |\la m_A |H^I_{\alpha \beta}| m_A'\ra|/N_A^2 )$
 for the indices $\alpha$ and $\beta$.

 \subsection{The case of quite weak interaction}
 \label{sect-swi}

 Let us first recall two properties proved in the canonical-typicality approach for large environments:

 \noindent P1. $\rho^S \simeq \rho^S_{\delta E}$  \cite{Pepes06}.

 \noindent P2. In the basis $|E^S_\alpha\ra $, $\rho^S_d$ has negligibly small off-diagonal elements
 and has diagonal elements proportional to $N^{(\E)}_\alpha$ \cite{Gold06}.
 The subspaces $\HH^{(\E)}_\alpha$ are assumed to have dimensions much larger than that of
 the Hilbert space of the system $S$.
 As usually done, the environment is assumed to have a density of states with exponential dependence
 on energy \cite{LL-SP}, then, $\rho^S_d$ has the usual canonical form.

 When the interaction is sufficiently weak such that
 the eigenstates $|E_\eta\ra $ are approximately equal to $|E^{S}_\alpha E^\E_i\ra$,
 one has $\HH_{\delta E} \simeq \HH_d$.
 Then, the above mentioned two properties P1 and P2 imply that
 $\rho^S $ has the usual canonical form.
% if the environment has a density of states with exponential dependence on energy.

 The above approach meets a difficulty when dealing with interaction with strength of practical interest,
 due to the exponential growth of the dimension of the Hilbert space
 and the at-most polynomial growth of the energy with increasing size of the environment,
 which implies exponential denseness of the spectrum of $H^S+H^\E$
 and deviation of $ \HH_d$ from $\HH_{\delta E}$.
 Recently, it was shown that the difficulty can be overcome by a perturbation approach
 and the usual canonical distribution is still obtained \cite{RGE12}.
 While, it is unclear how to use this method to solve the previously-mentioned problem
 of interest in this paper for non-extremely weak interaction.

 Therefore, we employ an alternative method, based on analysis in elements of the Hamiltonians.
% For this purpose, we consider an environment $\E$ with a division mentioned previously,
% namely, $\E = A + B$, where $A$ is much smaller than $B$ and $H^{I}$ couples $S$ to $A$ only.
 Below, we show that, when the interaction is sufficiently weak
 (but not so weak such that the interaction strength is not of practical interest),
 $\rho^S_{\delta E}\simeq\rho^S_d$, then, P1 and P2 imply that $\rho^S $ has
 approximately the usual canonical form.

 Of particular interest in our analysis are off-diagonal elements of the total
 Hamiltonian $H$ in the basis $|E^{S}_\alpha E^\E_i\ra$, which are given by
 $H^I_{\alpha i,\beta j}=(H^I_{\alpha\beta})_{ij}$ [Eq.(\ref{HIab2})].
 After some analysis (see appendix \ref{app1}), we find that
 $(H^I_{\alpha\beta})_{ij}$ have quite different properties depending on whether $i=j$ or not,
 specifically,
  \begin{subequations}
 \begin{eqnarray}\label{HIii}
 (H^I_{\alpha\beta})_{ii}  \simeq h^{\rm dia}_{\alpha \beta}, \hspace{3cm}
 \\ \label{HIij1} (H^I_{\alpha\beta})_{ij} \sim  N_A^{3/2} / N_\E^{1/2}  \ \  \ \ \text{for} \  i \ne j,
 \\ |(H^I_{\alpha\beta})_{ij}| \lesssim h N_A^{3/2} / N_\E^{1/2} \ \ \ \text{for} \  i \ne j,
 \label{HIij2} \end{eqnarray}
  \end{subequations}
 These relations imply that, in the general case with non-zero $h^{\rm dia}_{\alpha \beta}$, one has
 \begin{equation} \label{Hene}
 |(H^I_{\alpha\beta})_{ii}| \gg |(H^I_{\alpha\beta})_{ij}| \  \text{of} \  i\ne j,
 \end{equation}
 due to the largeness of the environment.
 Below, we discuss this general case of  non-zero $h^{\rm dia}_{\alpha \beta}$ \cite{foot-hdia}.

 For sufficiently small $\epsilon$, the estimate (\ref{HIii}) shows that $|(H^I_{\alpha\beta})_{ii}|$
 can be much smaller than the corresponding unperturbed-level spacings,
 $E^S_\alpha - E^S_\beta$ (the system $S$ having a non-degenerate spectrum).
 In this case,
 the influence of the elements $H^I_{\alpha i,\beta i}$ in the wave function of $|E_\eta\ra $
 in the basis $|E^S_\alpha E^\E_i\ra$ can be neglected.
 On the other hand, nearest-level spacings are on average approximately proportional to $1/N_\E$,
 while $(H^I_{\alpha\beta})_{ij}\sim 1/N_\E^{1/2}$ for $i\ne j$ [see (\ref{HIij1})],
 hence, as long as the environment is sufficiently large,
 couplings between nearest unperturbed levels are usually much larger than their spacings.
 Therefore, for whatever small but fixed $\epsilon$, each $|E_\eta\ra $ has significant expansions
 in many basis states $|E^S_\alpha E^\E_i\ra$, as a result, $\HH_{\delta E} \simeq \HH_d$ can not hold.

 To go further, we need to get an estimate to the width of the wave function of $|E_\eta\ra $ in
 the basis $|E^S_\alpha E^\E_i\ra$.
 As discussed above, in this weak-coupling case, the terms $(H^I_{\alpha\beta})_{ii}$ can be neglected.
 To get the estimate, we employ a first-order perturbation-theory treatment.
 The basic idea is that perturbation theory is applicable to components in those
 basis states whose unperturbed energies are sufficiently far from the eigenenergy $E_\eta$
 \cite{FGGK94,WIC98}, hence, perturbatively computing population of $|E_\eta\ra $ in all far-lying
 unperturbed basis $|E^S_\alpha E^\E_i\ra$, one may find an upper border for the width of
 the main body of the wave function.
 The result is that the main body of the wave function lies within a narrow energy window around $E_\eta$,
 with width $\delta e \propto 1/\Delta E$ (see appendix \ref{app2}).
 Here, $\Delta E$ is the scope of the eigenenergies of the total system, such that
 the averaged density of states is around $N_S N_\E / \Delta E$.
 In the large-environment limit, $\Delta E$ goes to infinity, hence, the width $\delta e$ shrinks to zero.

 Now, we specify the exact content of the condition (iv) discussed below Eq.(\ref{rho-c}), that is,
 we assume that the width $\delta E$ is fixed, or decreases sufficiently slowly with increasing
 size of the environment,
 such that $\delta e/ \delta E$ goes to zero in the large-environment limit.
 As shown in Appendix \ref{app3}, under this condition, $\rho^S_{\delta E}$ has properties similar to
 those of $\rho^S_d$ discussed above in P2, hence, has the usual canonical form.
 Then, according to P1, $\rho^S $ also has the usual canonical form.

 Finally, we observe that the above-used strong version of the second condition,
 i.e., the assumption of $\E$ being a quantum chaotic system,
 can in fact be loosed for the purpose of getting the above-discussed estimates to the elements
 $(H^I_{\alpha\beta})_{ij}$.
 Indeed, as shown in App.\ref{app4}, similar estimates to $(H^I_{\alpha\beta})_{ij}$ can be gotten under
 the following assumption about the environment, namely,
 \begin{itemize}
  \item $ G^{ii}_{mm}$ of a given $|E^\E_i\ra$ does not change much with $m$ and
 $|G^{ij}_{mm'}|$ with $j\ne i$ does not change much with $m$, $m'$, $i$ and $j$.
 \end{itemize}
 Obviously, this assumption is valid for the above-considered chaotic environment, hence, we call
 this assumption a weak version of the second condition.
 Then, following the same arguments as given above, one also finds a canonical form of $\rho^S$.

 Furthermore, it may happen that the above-discussed weak version of the second condition is satisfied
 only within a part of the spectrum of an environment of interest.
 For example, if the environment has a sufficiently chaotic motion in the
 middle energy region, while has a regular motion with integrable feature in the low energy region,
 then, the above-used approach is applicable only within the middle energy region.
 It is straightforward to extend discussions given in App.\ref{app4} to this case,
 where the relevant change will be that $N_\E$ in the relation (\ref{Hnd}),
 so in relations (\ref{HIij1}) and (\ref{HIij2}),
 is replaced by the number of the levels in the part of the spectrum of interest.

 \subsection{The case of relatively weak interaction}
 \label{sect-rwi}

 We use $\epsilon_{\rm ew}$ to indicate a border of $\epsilon$, beyond which
 some elements $(H^I_{\alpha\beta})_{ii}$ become non-negligible
 compared with the corresponding level spacings.
 The case of $\epsilon < \epsilon_{\rm ew}$ having been discussed in the previous section,
 in this section we discuss the case of $\epsilon > \epsilon_{\rm ew}$.
 In this case, arguments given in App.\ref{app2} and \ref{app3}, to show the canonical form of
 $\rho^S_{\delta E}$ in the basis $|E^S_\alpha \ra$,  do not apply.
 In fact, here, a total state $|E_\eta\ra$ may have comparable populations in
 related basis states $|E^S_\alpha E^\E_i\ra$ and $|E^S_\beta E^\E_i\ra$,
 as a consequence, the width of the main body of its wave function does not shrink to zero in
 the large-environment limit.

 A main observation is that the renormalization of the self and interaction Hamiltonians given in
 Eq.(\ref{reHS}) can significantly reduce the elements $(\ww H^I_{\ww\alpha \ww\beta})_{ii}$
 in the renormalized basis $|E_{\ww\alpha}\ra$.
 When the reduction is sufficiently large, arguments given in App.\ref{app2} and \ref{app3}
 will become applicable in the renormalized basis.
 To show this, below we discuss two situations separately, related to validity of the so-called
 `eigenstate thermalization hypothesis' (ETH) \cite{Deuts91,Sredn94}.
 The ETH, confirmed by direct numerical simulations \cite{Rigol08,DSFVJ12},
 effectively states that a few-body observable (e.g., $H^{I}_{\ww\alpha \ww\beta}$ here)
 has almost the same expectation values for energy eigenstates within an appropriate energy window
 of a large, interacting many-body system.
 We use $\HH^{(\E)}_{\rm ETH}$ to indicate a subspace spanned by eigenstates $|E^\E_i\ra $
 lying in an energy window of the environment, for which the ETH is applicable \cite{foot-ETH}.

 Let us first discuss the case that there exists a subspace $\HH^{(\E)}_{\rm ETH}$,
 which is sufficiently large to include all the subspaces $\HH^{(\E)}_{\ww\alpha}$ defined in
 Sec.\ref{sect-mainr}.
 In this case, $\la H^{I}_{\ww\alpha \ww\beta}\ra_{\ww\gamma} \simeq \la H^{I}_{\ww\alpha \ww\beta}
 \ra_{\rm ETH}$, where $\la H^{I}_{\ww\alpha \ww\beta} \ra_{\rm ETH}$ indicates
 the average of the expectation value $\la E^\E_i|H^{I}_{\ww\alpha \ww\beta}|E^\E_i\ra$ over
 $|E^\E_i\ra \in \HH^{(\E)}_{\rm ETH}$.
 Using this property and the definition of $H^I_S$ in Eq.(\ref{HIS}),
 we get the following explicit expression of $H^I_S$,
 \be \label{HIS1} H^I_S \simeq \sum_{\ww\alpha \ww\beta} \la H^{I}_{\ww\alpha \ww\beta}\ra_{\rm ETH}
 |E_{\ww\alpha}\ra \la E_{\ww\beta} |
 = \la H^{I} \ra_{\rm ETH}, \ee
 where $\la H^{I} \ra_{\rm ETH}$ indicates the average of $\la E^\E_i|H^{I}|E^\E_i\ra$
 over $|E^\E_i\ra \in \HH^{(\E)}_{\rm ETH}$, which is an operator acting on the Hilbert space of the
 system $S$.
 Substituting the expression of $\WW H^I$ in Eq.(\ref{reHS}) into the definition
 $(\WW H^I_{\ww\alpha \ww\beta})_{ii} = \la E_{\ww\alpha} E^\E_i|\WW H^I|E_{\ww\beta} E^\E_i\ra$
 and making use of Eq.(\ref{HIS1}) and the ETH, we find almost zero $(\WW H^I_{\ww\alpha \ww\beta})_{ii}$,
 namely,
 \begin{eqnarray}\label{HTnon1}
 (\WW H^I_{\ww\alpha \ww\beta})_{ii}  = \la  E^\E_i|H^{I}_{\ww\alpha \ww\beta}| E^\E_i\ra -
 \la E_{\ww\alpha} |H^I_S|E_{\ww\beta} \ra \simeq 0
 \end{eqnarray}
 for all $|E^{(\E)}_i\ra \in \HH^{(\E)}_{\rm ETH}$.

 Due to the smallness of the elements $(\WW H^I_{\ww\alpha \ww\beta})_{ii}$ in Eq.(\ref{HTnon1}),
 arguments used in App.\ref{app2} and \ref{app3} are applicable for the renormalized Hamiltonians
 and in the renormalized basis.
 Hence, the system $S$ has approximately the canonical statistical description in Eq.(\ref{rho-c})
 in terms of the renormalized self-Hamiltonian.
% if the environment has a density of states with exponential dependence on energy.
 Note that, since Eq.(\ref{HTnon1}) holds even for not small $\epsilon$,
 in this case of sufficiently large $\HH^{(\E)}_{\rm ETH}$,
 Eq.(\ref{rho-c}) may hold even when the system-environment interaction is not weak.

 Next, we consider the case that no subspace $ \HH^{(\E)}_{\rm ETH}$ is sufficiently large
 to include all the subspaces $\HH^{(\E)}_{\ww\alpha}$.
 Since we have assumed validity of the microcanonical-ensemble description for the total system
 within a narrow energy window $\delta E$,
 it is reasonable to assume that the ETH is applicable within each subspace
 $\HH^{(\E)}_{\ww\alpha}$ with the same width of energy window (the system $S$ being small).
 Then, for $E_{ \ww\alpha} > E_{\ww \beta}$,
 making use of Eq.(\ref{HIS}) and the definition of $\WW H^I$, we get
 \begin{eqnarray}\label{HTnon3}
 (\WW H^I_{\ww\alpha \ww\beta})_{ii}  \simeq
 \left \{
 \begin{array}{ccc}
 0 \ \ \ & \text{for} &  |E^\E_{i}\ra \in \HH^{(\E)}_{\ww\beta}
 \\ \la H^{I}_{\ww \alpha \ww\beta} \ra_{\ww\alpha} -
 \la H^{I}_{ \ww\alpha \ww\beta} \ra_{\ww\beta} \ \ \ & \text{for} &
 |E^\E_{i}\ra \in \HH^{(\E)}_{\ww\alpha}.
 \end{array}
 \right .
 \end{eqnarray}
 Since the dimension of $\HH^{(\E)}_{\ww\beta}$ is larger than that of $\HH^{(\E)}_{\ww\alpha}$
 for $E^S_{ \ww\alpha} > E^S_{\ww \beta}$, Eq.(\ref{HTnon3}) shows that
 most of the elements $(\WW H^I_{\ww\alpha \ww\beta})_{ii}$ are negligibly small.
 Meanwhile, since $(H^{I}_{\alpha \beta})_{ii} \simeq \la H^{I}_{\alpha \beta}
 \ra_{\gamma}$ for $|E^\E_{i}\ra \in \HH^{(\E)}_{\gamma}$ if the ETH is applicable within
 $\HH^{(\E)}_{\gamma}$, in general, the rest of the elements $(\WW H^I_{\ww\alpha \ww\beta})_{ii}$
 (the second line on the right hand side of Eq.(\ref{HTnon3}))
 are also reduced compared with $(H^{I}_{\alpha \beta})_{ii}$.
 It is easy to see that similar results also hold for $E^S_{ \ww\alpha} < E^S_{\ww \beta}$.
 To summarize, compared with $(H^{I}_{\alpha\beta})_{ii}$ in the unrenormalized case,
 the elements $(\WW H^I_{\ww\alpha \ww\beta})_{ii}$ are considerably reduced.

 As discussed in the previous section, for $\epsilon< \epsilon_{\rm ew}$, $(H^{I}_{\alpha\beta})_{ii}$
 are small enough to guarantee validity of the arguments used in App.\ref{app2} and \ref{app3},
 which predicts a canonical form of $\rho^S_{\delta E}$.
 The above-discussed reduction of $(\WW H^I_{\ww\alpha \ww\beta})_{ii}$
 implies that, at least for $\epsilon$ not too much above $\epsilon_{\rm ew}$,
 $(\WW H^I_{\ww\alpha \ww\beta})_{ii}$ should be small enough to guarantee
 validity of the arguments used in App.\ref{app2} and \ref{app3} for the renormalized Hamiltonians
 and in the renormalized basis.
 In this case, $\rho^S_{\delta E}$ has a canonical form in terms of the renormalized self-Hamiltonian,
 hence, Eq.(\ref{rho-c}) holds. This completes our proof.

 \section{Conclusion and discussions}

 In this paper, we have studied statistical description of a small quantum system $S$ in
 relatively weak interaction with a huge environment $\E$.
 It is shown that the system $S$
 has a canonical statistical description given by a renormalized self-Hamiltonian,
 which appropriately takes into account the influence of the system-environment interaction,
 if the following four conditions are satisfied:

 \noindent
 (i) The environment $\E$ can be divided into two parts $A$ and $B$,
 such that the interaction Hamiltonian $H^{I}$ couples $S$ to $A$ only,
 where $A$ is a small system and $B$ is much larger than $A$.
 (ii) Eigenstates of the environment $\E$ within a sufficiently large region of its spectrum
  have the property that
 $ G^{ii}_{mm}$ of a given $|E^\E_i\ra$ does not change much with $m$ and
 $|G^{ij}_{mm'}|$ of $j\ne i$ does not change much with $m$, $m'$, $i$ and $j$.
 (iii) The system-environment interaction is relatively weak,
 in the case that the ETH is applicable within each subspace $\HH^{(\E)}_{\ww\alpha}$ separately;
 while, there is no restriction to the interaction strength,
 if the ETH is applicable within the direct sum of the subspaces $\HH^{(\E)}_{\ww\alpha}$.
 (iv) The width $\delta E$ is fixed, or decreases sufficiently slowly
 such that $\delta e/ \delta E$ goes to zero in the large-environment limit.

 Some remarks concerning the above conditions may be helpful:
 (a) Concerning the above first condition,
 the division of $\E$ into $A$ and $B$ is not a too strict restriction in interaction,
 since it is usually satisfied in physical models of practical interest.
 While, the first condition may be unsatisfied, if the system-environment interaction
 has a long-range feature.
 (b) The above second condition is satisfied by an environment which is a quantum chaotic system
 that can be described by the random matrix theory.
 However, the exact relation between the above-discussed property of $ G^{ij}_{mm'}$
 stated in the second condition
 and integrability/non-integrability of the environment needs further investigation,
 because  many-body effects may also play important role here.
 (c) In the case that the ETH is completely inapplicable for the environment, discussions given
 in Sec.\ref{sect-rwi} do not apply, hence, it is not clear whether Eq.(\ref{rho-c}) is valid
 for relatively weak interaction in this case.

 It may be also helpful to give some further discussions for the example of spins mentioned in
 the section of introduction.
 (1) Models like Ising model with nearest-neighbor coupling satisfy the above first condition
 about separation of $\E$ into $A$ and $B$;
 here, $A$ is composed of spins that are the nearest neighbors of the spin $S$.
 Discussions given in this paper are not valid for the case that the spin $S$ is directly coupled
 to all the spins in its environment.
 (2) In a model like Ising model, when the coupling among the spins within the environment are
 not weak, it is usually reasonable to expect that the above second condition may be satisfied.

 Before concluding this paper, we give a brief discussion for possible experimental test in effects of
 some non-mean-field part of the renormalized Hamiltonian, e.g.,
 the term $\Delta H^{S}$ in Eq.(\ref{HSfull}) in the case of $H^I = \epsilon J^S J^A$.
 The term $\Delta H^{S}$ may cause notable effects, when the averaged expectation values of $J^A$
 in different energy regions $[E-E_{\ww\alpha},E-E_{\ww\alpha} +\delta E]$ have
 notable difference.
 For example, this may happen for an environment lying in a state in the vicinity of a
 (quantum) phase transition.
 In particular, if the environment has localized states
 and extended states separated by certain energy level, then,
 $J^A$ may have different expectation values below and above the energy level,
 which may have experimentally testable effects.
 Another possibility lies in that, for a relatively small environment (still much larger than the system $S$),
 it may be relatively easy for the averaged expectation value of $J^A$ to have notable
 variation with the energy $E_{\ww\alpha}$.

\acknowledgments
 We acknowledge partial financial support from the Natural Science Foundation of China under Grant
 No.10975123, the National Fundamental Research Programme of China,
 and `Boshidian' Foundation of the Ministry of Education of China.
%  Grant No.2007CB925200.

\appendix

 \section{Estimates to $(H^I_{\alpha \beta})_{ij}$ (I)}
 \label{app1}

 In this appendix, we give estimates to the elements $(H^I_{\alpha \beta})_{ij}
 = \la E^\E_{i} | H^I_{\alpha \beta} |E^\E_{j}\ra$, under the assumption
 that the environment can be regarded as
 a quantum chaotic system that can be described by the random matrix theory.
 Since $H^I$ couples the system $S$ to $A$ only, $H^I_{\alpha\beta}$ is in fact an operator
 acting on the Hilbert space of the system $A$ and its elements can be written as
 \be (H^I_{\alpha \beta})_{ij}
 = \sum_{m m'} \la m_A |H^I_{\alpha \beta}| m_A'\ra  G^{ij}_{mm'},
 \label{EHIE1} \ee
 where $G^{ij}_{mm'}$ is defined in Eq.(\ref{G}).
 Using the expansion in Eq.(\ref{Ei-expan}), we get
 \begin{equation}\label{Gex}
 G^{ij}_{mm'} = \sum_q \left (C^i_{mq}\right )^* C^j_{m'q}.
 \end{equation}

 Normalization of $|E^\E_i\ra $ requires that
 \begin{equation}\label{nC}
 \sum_{m q} |C^i_{mq}|^2 =1.
 \end{equation}
 As assumed, the real and imaginary parts of $C^i_{mq}$ can be regarded as random numbers
 with mean zero and the same variance,
 hence, Eq.(\ref{nC}) implies that most of the values of $|C^i_{mq}|$ are around $ N_\E^{-1/2}$.
 Then, noticing the largeness of $N_B$ and that $N_\E = N_A N_B$, making use of Eq.(\ref{Gex})
 and the randomness of $C^i_{mq}$,  we get
 \begin{equation}\label{}
 |G^{ij}_{mm'}| \simeq (N_\E N_A)^{-1/2}  \ \ \ \ \text{for} \ i\ne j \ \text{or} \ m\ne m' .
 \end{equation}
 For $i=j$ and $m=m'$, we have
 \begin{equation}\label{}
 G^{ii}_{mm} \simeq 1/N_A.
 \end{equation}

 Substituting the above-obtained estimates to $G$ into Eq.(\ref{EHIE1}) and using the approximation
 $(1- N_A^{1/2} N_\E^{-1/2}) \simeq 1$, we get
 \begin{equation}\label{Hmc1}
 (H^I_{\alpha \beta})_{ii} \simeq h^{\rm dia}_{\alpha \beta} + \Delta h,
 \end{equation}
 where $h^{\rm dia}_{\alpha \beta}$ is defined in Eq.(\ref{hdia})
% \begin{eqnarray}
%   h^{\rm dia}_{\alpha \beta} = \frac{1}{N_A} \sum_m \la m_A |H^I_{\alpha \beta}| m_A\ra
% \end{eqnarray}
 and $\Delta h$ satisfies $|\Delta h| \lesssim h_{\alpha \beta} N_A^{3/2} N_\E^{-1/2}$.
 Here,
 \begin{eqnarray}
  \\ h_{\alpha \beta} \equiv \frac 1{N_A^2}  \sum_{m m'} |\la m_A |H^I_{\alpha \beta}| m_A'\ra|,
 \end{eqnarray}
 i.e., the average of the absolute values of the elements of $H^I_{\alpha \beta}$.
 Similarly, we find
 \begin{equation}\label{Hnd1}
 |(H^I_{\alpha \beta})_{ij}| \lesssim h_{\alpha \beta} N_A^{3/2} N_\E^{-1/2}
 \lesssim h N_A^{3/2} N_\E^{-1/2} \ \ \text{for} \ i\ne j,
 \end{equation}
 where $h$, defined below Eq.(\ref{hdia}), is just the maximum of $h_{\alpha \beta}$.

 \section{Width of $|E_\eta\ra $ in basis $|E^0_k\ra $ for sufficiently small $\epsilon$}
 \label{app2}

 For brevity, we use $|E^0_k\ra $ to denote the basis states $|E^S_\alpha E^\E_i\ra $
 ($E^0_k = E^S_\alpha + E^\E_i$) with increasing energy order,
 and use $C_{\eta k}$ to denote the components of an eigenstate
 $|E_\eta\ra $ in the basis $|E^0_k\ra$, namely, $C_{\eta k} = \la E^0_k | E_\eta\ra $.
 The main body of the wave function of $|E_\eta\ra $ in the basis $|E^0_k\ra $
 lies within a region denoted by $ [k_1,k_2]$, outside which the total population of $|E_\eta\ra$
 is equal to some small positive number $\epsilon_p$, namely, $P_{k_1 k_2} = \epsilon_p$,
 where
 \begin{equation}\label{P12}
 P_{k_1 k_2} = \sum_{k\in Q} |C_{\eta k}|^2.
 \end{equation}
 Here and after, $Q$ denotes the set of the label $k$ lying outside the region $[k_1,k_2]$,
 i.e., $Q=\{ k: k<k_1 \ \text{or} \ k>k_2 \}$.

 Below in this appendix,
 using a first-order perturbation theory and results of App.\ref{app1},
 we derive an estimate to the width $\delta e = E^0_{k_2} - E^0_{k_1}$, which gives
 an upper border for the width of the main body of the wave function of $|E_\eta\ra$.
 Using $|E^0_{k_0}\ra = |E^S_{\alpha_0} E^\E_{i_0}\ra $ to denote
 the basis state whose energy is the closest to the total energy $E_\eta$,
 the first-order perturbation theory gives the following prediction for the population $P_{k_1 k_2}$,
 which we denote by $P_{k_1 k_2}^{(1)}$,
 \begin{equation}\label{P}
 P_{k_1k_2}^{(1)} = \sum_{k \in Q} \left | \frac{(H^I_{\alpha_0 \beta })_{i_0 j}}
 {E^0_{k_0} - E^0_k} \right |^2 \ \  \text{with} \ |E^0_k\ra =|E^S_\beta E^\E_j\ra .
 \end{equation}
 Note that $k_0$ lies in the region $[k_1,k_2]$, hence, not belongs to $ Q$.

 Then, substituting the relation (\ref{Hnd1}) for $j\ne i_0$ into Eq.(\ref{P}), we get
 \begin{equation} \label{P1}
 P_{k_1k_2}^{(1)} \lesssim \frac{h^2 N_A^3}{N_\E }
 {\sum_{k\in Q}}' \left | \frac{1}{E^0_{k_0} - E^0_k} \right |^2 + p_d,
 \end{equation}
 where the prime means that the summation is over those $k\in Q$ for which $j\ne i_0$ and
 $p_d$ is the contribution from $k\in Q$ with $j= i_0$, namely,
 \begin{equation}\label{pd}
 p_d \equiv \sum_{k\in Q \& j=i_0} \left | \frac{(H^I_{\alpha_0 \beta })_{i_0 j}}
 {E^0_{k_0} - E^0_k} \right |^2 .
 \end{equation}
 Noticing that the label $k$ represents the pair $(\beta,j)$ and the summation over
 ``$k\in Q \& j=i_0$'' is usually given by a summation over some of the labels $\beta \ne
 \alpha_0$, we get
 \begin{equation}\label{Pa}
 p_d \le  p_{\alpha_0 i_0},
 \end{equation}
 where
 \begin{equation}\label{}
 p_{\alpha_0 i_0} = \sum_{\beta (\ne \alpha_0) } \left |
 \frac{(H^I_{\alpha_0 \beta })_{i_0 i_0}} {E^S_{\alpha_0} - E^S_\beta} \right |^2.
 \end{equation}
 Using the estimate in (\ref{Hmc1}) and neglecting $\Delta h$, we find
 \begin{equation}\label{Pai}
 p_{\alpha i} \lesssim (h_{d})^2 q_\alpha \ \ \  \text{with}
 \ q_\alpha= \sum_{\beta (\ne \alpha) } \left |
  {E^S_{\alpha} - E^S_\beta} \right |^{-2}.
 \end{equation}
 Here, $h_{d}$ is the maximum of $|h^{\rm dia}_{\alpha \beta}|$.

 When $\epsilon $ is sufficiently small,
 since $h_d$ is proportional to $\epsilon$ and $ q_\alpha$ is independent of $\epsilon$,
 one has $p_{\alpha i} \ll \epsilon_p$ for a given $\epsilon_p$.
 Then, due to the relation (\ref{Pa}),
 the term $p_d$ on the right hand side of (\ref{P1}) can be neglected.

 To go further, since the number of $k$ for which $j=i_0$ is much smaller than that
 for which $j\ne i_0$, the prime over the summation on the right hand side
 of (\ref{P1}) can be taken away.
 Then, replacing the summation  by
 an integration over $E^0$, namely, by $\int dE^0 \rho(E^0)$, where $\rho(E^0)$ is the density of states,
 we get
 \begin{equation} \label{P2}
 P_{k_1k_2}^{(1)} \lesssim \frac{h^2 N_A^3}{N_\E }
 \int ' dE^0  \frac{\rho(E^0)}{|E^0_{k_0} - E^0|^2},
 \end{equation}
 where the prime indicates the integration domain $(-\infty ,E^0_{k_1}] \bigcup [E^0_{k_2},\infty )$.

 To have a rough estimate, we make a further approximation that $\rho(E^0)$ in (\ref{P2})
 can be approximated by its average value within an energy region with width $\Delta E$, which
 includes most of the levels of the total system, that is, $\rho(E^0) \simeq N_\E N_S / \Delta E $.
 Then, performing the integration on the right hand side of  (\ref{P2}), we have
 \begin{equation} \label{P3}
 P_{k_1k_2}^{(1)} \lesssim \frac{ h^2 N_A^3 N_S}{\Delta E}
 \left ( \frac{1}{|E^0_{k_0} - E^0_{k_1}|} + \frac{1}{|E^0_{k_0} - E^0_{k_2}|} \right ) .
 \end{equation}
 Replacing $P_{k_1k_2}^{(1)}$ by $\epsilon_p$, the relation (\ref{P3}) gives the following estimate to
 $\delta e = E^0_{k_2} - E^0_{k_1}$,
 \begin{equation} \label{de1}
 \delta e^{(1)}  \lesssim \frac{ 4h^2 N_A^3 N_S}{\epsilon_p \Delta E}.
 \end{equation}
 In the limit that the $B$ part of the environment becomes infinitely large,
 $\Delta E$ goes to infinity, while, both $N_S$ and $N_A$ remain finite.
 Therefore, for each fixed $\epsilon_p$, $\delta e^{(1)} \to 0$,
 that is, the energy width of $|E_\eta\ra $ in the basis $|E^0_k\ra $ shrinks to zero,
  in the limit of large environment.

 Finally, we give a remark on the validity of using perturbation theory to give the above estimates.
 In fact, as shown in Refs.\cite{WIC98,GBW},
 for $E^0_{k_1}$ and $E^0_{k_2}$ sufficiently far away from $E_\eta$,
 the components $C_{\eta k}$ of $k \in Q$ can  be
 expressed in convergent perturbation expansions in terms of $C_{\eta k}$ of $k\in [k_1,k_2]$
 .
% and, furthermore, the closest $k_1$ and $k_2$ can give an estimate for the scope of the main
% body of the wave function \cite{GBW}.

 \section{Canonical form of $\rho^S_{\delta E}$ for sufficiently small $\delta e$}
 \label{app3}

 In this section of appendix, we show that $\rho^S_{\delta E}$ has the usual canonical form,
 if $\delta E$ decreases sufficiently slowly with increasing size of the environment, such that
 $\delta e/\delta E$ goes to zero in the limit of large environment
 [cf. the inequality (\ref{de1})].

 Normalized typical vectors in the subspace $\HH_{\delta E}$ can be written as
\begin{equation}\label{aPt-de}
 |\Psi_{\rm ty}^{\delta E}\ra = \N_{\delta E}^{-1} \sum_{\eta \in \Gamma_{\delta E}} C_\eta |E_\eta\ra ,
\end{equation}
 where $\Gamma_{\delta E}$ is the set of the label $\eta $ for which
 $|E_\eta\ra \in \HH_{\delta E}$, namely, $\Gamma_{\delta E} \equiv \{ \eta:
 |E_\eta\ra \in \HH_{\delta E} \}$,
 and the real and imaginary parts of $C_\eta$ are independent real Gaussian random
 variables with mean zero and variance $1/2$.
 Let us expand the eigenstates $|E_\eta\ra $ in the basis $|E^S_\alpha E^\E_i\ra $,
 \begin{equation}\label{Ee-ex}
 |E_\eta\ra = \sum_{\alpha, i} f^\eta_{\alpha i} |E^S_\alpha E^\E_i\ra .
 \end{equation}
 Substituting Eq.(\ref{Ee-ex}) into Eq.(\ref{aPt-de}), we write $|\Psi_{\rm ty}^{\delta E}\ra$
 in a form similar to that of a typical vector in the subspace $\HH_d$, which has been used in
 Ref.\cite{Gold06} in deriving properties of $\rho^S_d$, namely,
 \begin{equation}\label{}
 |\Psi_{\rm ty}^{\delta E}\ra = \N_{\delta E}^{-1} \sum_{\alpha} |E^S_\alpha\ra |\Omega^\E_\alpha\ra ,
 \end{equation}
 where
 \begin{eqnarray} \label{Omega}
 |\Omega^\E_\alpha\ra =  \sum_{i} K_{\alpha i} | E^\E_i\ra ,
 \\ \label{K} K_{\alpha i} = \sum_{\eta\in \Gamma_{\delta E}}
 C_\eta f^\eta_{\alpha i}.
 \end{eqnarray}
 Noticing that $\N_{\delta E}^2 = \sum_{\eta \in \Gamma_{\delta E }} |C_\eta|^2 \simeq N_{\delta E}$,
 it is straightforward to verify that
 \begin{eqnarray} \label{rsab}
 (\rho^S_{\delta E})_{\alpha \beta} \simeq N_{\delta E}^{-1} \la \Omega^\E_\beta |\Omega^\E_\alpha \ra ,
 \\ \label{Osum}  \la \Omega^\E_\beta |\Omega^\E_\alpha \ra
 = \sum_i K_{\beta i}^* K_{\alpha i}.
 \end{eqnarray}

 Let us first discuss the case of $\alpha =\beta $, in which
 \begin{equation}\label{Oms}
 \la \Omega^\E_\alpha|\Omega^\E_\alpha\ra = \sum_i  \sum_{\eta\in \Gamma_{\delta E}}
 \sum_{\eta'\in \Gamma_{\delta E}}  C_\eta^* C_{\eta'} (f^\eta_{\alpha i})^*
  f^{\eta'}_{\alpha i}.
 \end{equation}
 The main contribution to the right hand side of Eq.(\ref{Oms}) comes from
 the diagonal terms with $\eta =\eta'$, hence,
 \begin{equation}\label{Omf}
 \la \Omega^\E_\alpha|\Omega^\E_\alpha\ra \simeq  \sum_i \sum_{\eta\in \Gamma_{\delta E}}
 |C_{\eta}|^2 |f^\eta_{\alpha i}|^2 .
 \end{equation}
 On the right hand side of Eq.(\ref{Omf}), with $\alpha $ fixed,
 we need to consider only those indices $i$,
 for each of which the basis state $|E^S_\alpha E^\E_i\ra $ lies within the main
 body of some state $|E_\eta\ra $ of $\eta\in \Gamma_{\delta E}$.
 That is, we need to consider only those $i$, for which the corresponding energies $E^\E_i$
 lie in the energy region $[E-E^S_\alpha-\delta e, E+\delta E -E^S_\alpha +\delta e]$.
 We use $N_{\alpha, \delta e}^{(\E)}$ to denote the number of levels $E^\E_i$ lying in this energy region.

 Now, we are ready to compute $(\rho^S_{\delta E})_{\alpha \alpha}$.
 Since $\delta e \ll \delta E$, for most of the indices $i$ mentioned in the previous paragraph, we have
 \begin{equation}\label{}
 \sum_{\eta\in \Gamma_{\delta E}} |f^\eta_{\alpha i}|^2 \simeq 1,
 \end{equation}
 which implies that
 \begin{equation}\label{}
 \sum_{\eta\in \Gamma_{\delta E}} |C_\eta|^2 |f^\eta_{\alpha i}|^2 \approx 1.
 \end{equation}
 Another result of $\delta e \ll \delta E$ is that $N_{\alpha, \delta e}^{(\E)} \approx N^{(\E)}_\alpha $.
 Then, Eq.(\ref{Omf}) gives $ \la \Omega^\E_\alpha|\Omega^\E_\alpha\ra \approx N^{(\E)}_\alpha $, hence,
 \begin{equation}\label{rho-dia}
 (\rho^S_{\delta E})_{\alpha \alpha} \approx N^{(\E)}_\alpha / N_{\delta E}.
 \end{equation}

 Next, we discuss the case of $\alpha \ne \beta$.
 It proves useful to explicitly express some properties related to the narrowness of
 the wave functions of $|E_\eta\ra $.
 We use $n_m$ to denote the average number of major components $f^\eta_{\alpha i}$ of
 $|E_\eta\ra $ in the basis $|E^S_\alpha E^\E_i\ra $.
% for the simplicity in discussion, we assume that other components can be neglected in our discussion.
 Thus, on average, $|f^\eta_{\alpha i}| \approx (n_m)^{-1/2}$.
 The major components lie within the main body of the corresponding wave function with width $\delta e$.
 In the limit of large environment, since $\delta e/\delta E$ goes to zero,
 $n_m/N_{\delta E}$ also goes to zero.

 Note that each state $|E^S_\alpha E^\E_i\ra $ also has on average $n_m$ major components, when
 $|E_\eta\ra $ is taken as the basis.
 Hence, the summation on the right hand side of Eq.(\ref{K})
 is effectively taken over $n_m$ major terms.
 Then, due to the randomness of the
 coefficients $C_\eta$ and the relation $|f^\eta_{\alpha i}| \approx (n_m)^{-1/2}$,
 we have $|K_{\alpha i}| \sim 1$.
 Furthermore, as discussed above, for a given $\alpha$, the number of the indices $i$
 needing consideration is about $N_{\alpha, \delta e}^{(\E)} \approx N^{(\E)}_\alpha $,
 hence, the number of non-negligible $K_{\alpha i}$ is also about $ N^{(\E)}_\alpha $.
 Meanwhile, the number of the random coefficients $C_\eta$ is $N_{\delta E}
 =\sum_\alpha N^{(\E)}_\alpha$.
 Therefore, $K_{\beta i}$ and $K_{\alpha i}$ with $\beta \ne \alpha$ in the summation in Eq.(\ref{Osum})
 can be treated as independent random numbers.
 Then, from Eqs.(\ref{rsab}) and (\ref{Osum}), we get for $\alpha \ne \beta $,
 \begin{equation}\label{rho-nd}
 |(\rho^S_{\delta E})_{\alpha \beta} | \lesssim (N_{\delta E})^{-1/2}.
 \end{equation}
 Equations (\ref{rho-dia}) and (\ref{rho-nd}) show that $\rho^S_{\delta E}$ has approximately
 the same properties as $\rho^S_d$ stated in P2 (see Sec.\ref{sect-swi}), hence,
 $\rho^S_{\delta E}$ also has the usual canonical form.

 \section{Estimates to $(H^I_{\alpha \beta})_{ij}$ (II)}
 \label{app4}

 In this appendix, we show that estimates, similar to those given in App.\ref{app1} for
 the elements $(H^I_{\alpha \beta})_{ij}$, can be obtained under the assumption
 that $ G^{ii}_{mm}$ of a given $|E^\E_i\ra$ does not change much with $m$ and
 $|G^{ij}_{mm'}|$ of $j\ne i$ does not change much with $m$, $m'$, $i$ and $j$.
 This assumption is weaker than the assumption that the environment is a quantum chaotic system
 that can be described by the random matrix theory.

 Let us first discuss the case of $i=j$.
 A major contribution to $(H^I_{\alpha \beta})_{ii}$ from the right hand side of Eq.(\ref{EHIE1})
 comes from the diagonal terms with $m=m'$, for which $G^{ii}_{mm} = \| \la m_A|E^\E_i\ra \|^2$.
 Since $ G^{ii}_{mm} $ does not change much with $m$ and
 \begin{equation}\label{}
 \sum_m G^{ii}_{mm} = \la E^\E_i | E^\E_i \ra =1,
 \end{equation}
 we have $G^{ii}_{mm} \simeq 1/N_A$.
 This gives the following estimate to the above-mentioned major contribution to
 $(H^I_{\alpha \beta})_{ii}$, namely,
 \begin{equation}\label{Hmc}
 \sum_{m} \la m_A |H^I_{\alpha \beta}| m_A\ra  G^{ii}_{mm} \simeq h^{\rm dia}_{\alpha \beta}.
\end{equation}
% where $h_{\rm dia}^{\alpha \beta}$ indicates the average value of $\la m_A |H^I_{\alpha \beta}| m_A\ra$,
% \be h_{\rm dia}^{\alpha \beta} \equiv \frac{1}{N_A} \sum_m \la m_A |H^I_{\alpha \beta}| m_A\ra .
% \ee
 Furthermore, since $|G^{ii}_{mm'}|$ with $m\ne m'$ is usually not larger than $G^{ii}_{mm}$,
 Eq.(\ref{EHIE1}) and the above result $G^{ii}_{mm} \simeq 1/N_A$
 give the following estimate for an upper border of $|(H^I_{\alpha \beta})_{ii}|$, namely,
 \begin{equation}\label{Hii}
 |(H^I_{\alpha \beta})_{ii}| \lesssim N_A h.
 \end{equation}

 To get an estimate to $|G^{ij}_{mm'}|$ with $j\ne i$,  we note that
 \begin{eqnarray}
 \nonumber \sum_{j} |G^{ij}_{mm'}|^2 = \sum_{j} \la E^\E_{i} |m_A \ra \la m_A'|
 E^\E_{j} \ra \la E^\E_{j} |m_A' \ra \la m_A| E^\E_{i} \ra
 \\ = \la E^\E_{i} |m_A \ra  \la m_A| E^\E_{i} \ra  \simeq \frac 1{N_A}, \ \ \
 \label{aG1}
 \end{eqnarray}
 where the identity $\sum_j |E^\E_{j} \ra \la E^\E_{j} | =1$ has been used.
 As assumed, $|G^{ij}_{mm'}|$ of $j\ne i$ does not change much with $m$, $m'$, $i$ and $j$,
 hence, Eq.(\ref{aG1}) implies that, in most cases,
 \begin{equation}\label{}
 |G^{ij}_{mm'}| \simeq (N_\E N_A)^{-1/2} \ \ \text{for} \ i\ne j .
 \end{equation}
 Using this estimate and Eq.(\ref{EHIE1}), we get
 \begin{equation}\label{Hnd}
 |(H^I_{\alpha \beta})_{ij}| \lesssim h N_A^{3/2} N_\E^{-1/2} \ \ \text{for} \ i\ne j.
 \end{equation}

% \nonindent \textbf{Supplementary Information} is linked to the online version


\begin{thebibliography}{99}
\bibitem{foot-HIS}
 When the interaction is weak, one may use perturbation theory to calculate the renormalized term
 $H^I_S$.
 For example, one may use the eigenbasis of $H^{S}$ to replace the basis used on the
 right hand side of Eq.(\ref{HIS})  to get a first order approximation for the renormalized
 self-Hamiltonian $\WW H^S$, then, proceed with similar procedure iteratively.
\bibitem{foot-hdia} Note that $H^I_{\alpha\beta}$ defined in Eq.(\ref{Hab})
 with $\alpha \ne \beta$ are not changed
 by adding a multiple of the identity to the interaction Hamiltonian $ H^I$ and similar for
 $h^{\rm dia}_{\alpha \beta}$.
\bibitem{foot-ETH} It is of interest to note that, according to a mathematical theorem,
 namely, Levy's lemma, % \cite{MS01},
 typical vectors in $\HH^{(\E)}_{\rm ETH}$ (if it is sufficiently large)
 have almost the same expectation values for local operators.
 Hence, the ETH effectively assumes that the eigenvectors $|E^\E_i\ra$
 in $\HH^{(\E)}_{\rm ETH}$ can be regarded as typical vectors in the subspace.
\end{thebibliography}
 \end{document}